\documentclass
[prl,twocolumn,superscriptaddress,floatfix,showpacs,10pt]{revtex4}%
\usepackage{amsfonts}
\usepackage{amsmath}
\usepackage{amssymb}
\usepackage{graphicx}%
\begin{document}
\title{Manifestly Lorentz invariant chiral boson action}
\author{Paul K. Townsend}
\affiliation{Department of Applied Mathematics and Theoretical Physics, Centre for
Mathematical Sciences, University of Cambridge, Wilberforce Road, Cambridge,
CB3 0WA, UK}

\begin{titlepage}
\vfill
\end{titlepage}

\begin{abstract}

A manifestly Lorentz invariant action is found for the Floreanini-Jackiw chiral boson. The method involves a novel chiral reduction of the 
phase-space action for a string, and can be adapted to describe chiral bosons on the heterotic string worldsheet.  A similar manifestly
Lorentz invariant action is found for an entire class of  conformal chiral $2k$-form electrodynamics in $(4k+2)$ dimensions which 
includes the Floreanini-Jackiw theory as the $k=0$ case. 

\end{abstract}

\maketitle

\setcounter{equation}{0}

Maxwell's  conformal-invariant  equations for electrodynamics have a natural generalization to $n$-form electrodynamics in  a Minkowski spacetime of $2(n+1)$ dimensions, and these equations are derivable from a Lorentz-invariant action quadratic in the gauge-invariant $(n+1)$-form field strength (see e.g.  \cite{Henneaux:1986ht}).  For $n=2k$ one may consistently impose a self-duality condition on this $(2k+1)$-form; the resulting parity-violating, but still conformal invariant,  equations are those of  ``chiral $2k$-form electrodynamics'' (see e.g. \cite{Henneaux:1988gg}).  However, the self-duality condition makes it difficult to find a manifestly Lorentz-invariant action \cite{Marcus:1982yu}, which greatly complicates any
attempt to include interactions.  Many solutions to this problem have been proposed, each with some feature that could be considered undesirable; the most relevant here is that of Pasti et al. \cite{Pasti:1996vs} and the most recent is that of Sen \cite{Sen:2019qit} to whom we defer for references to other proposals. 

The focus here will be on a class of ``chiral $2k$-form electrodynamics'' that includes, as the $k=0$ case, the chiral boson theory of 
Floreanini and Jackiw \cite{Floreanini:1987as}. This is a conformally invariant free field theory for a scalar field $\varphi$ in a 2-dimensional Minkowski spacetime. For 
time-space coordinates $(t,\sigma)$ the Floreanini -Jackiw (FJ) action is 
\begin{equation}\label{FJact2}
S_{FJ}[\varphi]= \int\! dt \! \int \! d\sigma\,  (\dot\varphi - \varphi')\varphi'\, , 
\end{equation}
where the overdot indicates a partial time derivative and a prime indicates a derivative with respect to $\sigma$. The corresponding field equation is 
\begin{equation}\label{FJeq}
\partial _- \varphi' =0\,  \quad \left[\partial_\pm = \frac12(\partial_t \pm \partial_\sigma) \right]\, . 
\end{equation}
This is similar to the manifestly Lorentz invariant free chiral boson equation $\partial_- \varphi=0$, but with $\varphi'$ replacing the scalar field $\varphi$.
Neither the FJ action nor its field equation is manifestly Lorentz invariant but the first-order variation of the action 
under a Lorentz transformation is a surface term \cite{Floreanini:1987as}, and this implies Lorentz invariance of the equation (\ref{FJeq}); its general solution is 
\begin{equation}\label{gensol}
\varphi = \varphi_0(t)  + \Phi(\xi^+) \qquad (\xi^\pm = t\pm \sigma) \, ,  
\end{equation}
but the zero mode is unphysical because the FJ action has a (restricted) gauge invariance: $\varphi(\xi) \to \varphi(\xi) +a(t)$ for arbitrary function $a(t)$.  

One purpose of this paper is to show that the FJ action is a gauge-fixed version of a manifestly Lorentz invariant action constructed from fields that transform linearly under the Lorentz group. This `covariant' action can be interpreted  as a ``chiral dimensional reduction''   of the action for a string in a three-dimensional (3D) spacetime, and the 
method can be adapted to provide a covariant action for chiral bosons on the heterotic string.  

Another purpose of this paper is to show that the FJ chiral boson theory, in both its  original and covariant forms, is the special $k=0$ case of a class of conformal chiral $2k$-form electrodynamics 
in a $(4k+2)$-dimensional Minkowski spacetime. The $k=1$ case and its relation to the M5-brane \cite{Gibbons:2000ck,Townsend:2019ils} will be briefly reviewed towards the end of this paper before consideration of $k>1$. 

We begin with the 3D string of tension $T$. Let $\{X^m; m=0,1,2\}$ be Minkowski spacetime coordinates and let $(t,\sigma)$ now be arbitrary worldsheet coordinates.  
Prior to chiral reduction, the phase-space action is 
\begin{equation}\label{string}
S  = \int\! dt\int\! d\sigma \left\{ \dot X^m P_m - e {\cal H}_\perp - u{\cal H}_\parallel \right\}\, , 
\end{equation}
where the worldvolume fields $P_m$ are canonically conjugate to the maps $X^m$ from the worldsheet to spacetime,  and  
$(e,u)$ are Lagrange multipliers for phase-space constraints. The constraint functions are
\begin{eqnarray}
{\cal H}_\perp &=& \frac12\left[\eta^{mn}P_mP_n + T^2 \eta_{mn} (X^m)'(X^n)'\right] \, , \nonumber \\
 \qquad {\cal H}_\parallel &=& (X^m)'P_m \, . 
\end{eqnarray}
As is well known, these constraints are first class and their Poisson bracket (PB) algebra is the Lie algebra of the infinite-dimensional 2D conformal group. 
The gauge transformations generated by the constraints are on-shell equivalent to worldsheet diffeomorphisms. 

We shall now break the 3D Lorentz invariance to 2D Lorentz invariance 
by making the identification $X_2 \sim X_2 + 2\pi R$. On setting
\begin{equation}
X_2 = R\varphi \, , \qquad RP_2 = P_\varphi\, , 
\end{equation}
we get a phase-space action of the form 
\begin{equation}\label{string2}
S  = \int\! dt\int\! d\sigma \left\{ \dot X^\mu P_\mu + \dot\varphi P_\varphi - e {\cal H}_\perp - u{\cal H}_\parallel \right\}\, , 
\end{equation}
where $\mu=0,1$ and $\varphi \sim \varphi +2\pi$.  We shall suppose (in units for which $\hbar=1$) that 
\begin{equation}\label{TR21}
TR^2=1\, , 
\end{equation}
in which case the constraint functions can be written as 
\begin{eqnarray}\label{constraints}
{\cal H}_\perp &=& \frac12\left[P^2 + T \left(P_\varphi^2 + (\varphi')^2 \right) + T^2 (X')^2 \right] \, , \nonumber \\
 \qquad {\cal H}_\parallel &=& (X^\mu)'P_\mu + \varphi' P_\varphi \, ,  
\end{eqnarray}
where $P^2= \eta^{\mu\nu} P_\mu P_\nu$ and $(X')^2 = \eta_{\mu\nu} (X^\mu)'(X^\nu)'$.  The surviving 2D 
Lorentz group has only one generator, a Lorentz boost,  and its Noether charge is 
\begin{equation}\label{boost}
L= \int \! d\sigma \left(X^0 P^1 - X^1 P^0\right)\, . 
\end{equation}

To get some intuition into what has been done so far, we may impose the following Monge gauge conditions:  
\begin{equation}\label{Monge2}
X^0 =t, \quad X^1 = \sigma\, .
\end{equation}
In this gauge the constraints can be solved for $P_\mu$:
\begin{equation}\label{Pmu}
P_0= \mp\sqrt{\left[ T+ (\varphi')^2\right] \left[T+ P_\varphi^2\right]} \, , \qquad P_1 = - \varphi' P_\varphi \, . 
\end{equation}
Notice that the energy density $P^0$ has no $T\to\infty$ limit unless we subtract $T$ from it.  This may be achieved by 
making the following replacement in the constraints: 
\begin{equation}\label{replacement}
P_\mu \to  \mathbb{P}_\mu = P_\mu \mp T\varepsilon_{\mu\nu} (X^\nu)'\, . 
\end{equation}
This has no effect on the algebra of constraint functions but in the Monge gauge we now have
\begin{equation}
\mathbb{P}_0 = P_0 \mp T \, , \qquad \mathbb{P}_1 = P_1\, , 
\end{equation}
where the $P_\mu$ are as in (\ref{Pmu}). For the same sign choice as before we now have the Hamiltonian density
\begin{equation}
P^0 = \sqrt{\left[ T+ (\varphi')^2\right] \left[T+ P_\varphi^2\right]} -T\, .
\end{equation}
Notice that 
\begin{equation}
\lim_{T\to\infty} P^0 =  \frac12 \left[P_\varphi^2 + (\varphi')^2\right] \, .  
\end{equation}
Elimination of $P_\varphi$ now yields the standard Lorentz invariant free-field action for massless scalar field $\varphi$.

To get a chiral boson action from the string action (\ref{string2}) we must include an additional constraint: $P_2 \equiv TX_2'$.  Given the periodic identification of $X_2$ and the relation (\ref{TR21}), 
this constraint is equivalent to 
\begin{equation}\label{chi}
\chi(\sigma)\equiv 0 \, , \quad \chi = P_\varphi -\varphi' \, . 
\end{equation}
Expansion of $\chi(\sigma)$ on  a set of basis functions yields one zero mode (generator of  the $\varphi \to \varphi + a(t)$ gauge transformation of the FJ theory) and
a set of second-class constraints which would require, for some purposes, a replacement of Poisson brackets by Dirac brackets. 
This complication can be avoided by simply substituting $\varphi'$ for $P_\varphi$ in (\ref{string2}) to get the new action 
\begin{equation}\label{string22}
S  = \int\! dt\int\! d\sigma \left\{ \dot X^\mu P_\mu + \dot\varphi \varphi'  - e {\cal H}_\perp - u{\cal H}_\parallel \right\}\, , 
\end{equation}
where now
\begin{eqnarray}\label{Hamcons}
{\cal H}_\perp &=& \frac12 \left[\mathbb{P}^2 + 2T (\varphi')^2 + T^2 (X')^2\right]\, ,\nonumber \\
{\cal H}_\parallel &=& (X^\mu)'P_\mu + (\varphi')^2 \, . 
\end{eqnarray}
We have included here the replacement of (\ref{replacement}), which affects only ${\cal H}_\perp$. 
The canonical PB relations determined by this action are
\begin{eqnarray}\label{PBrels}
\left\{X^\mu(\sigma), P_\nu(\varsigma) \right\}_{PB} &=& \delta^\mu_\nu\,  \delta(\sigma-\varsigma)\, ,  \nonumber \\
\left\{\varphi(\sigma) , \varphi(\varsigma)\right\}_{PB} &=& -\frac12 \epsilon(\sigma-\varsigma)\, ,
\end{eqnarray}
where $\epsilon'(\sigma) = \delta(\sigma)$; the second line is also the Dirac bracket relation required for consistency with the chirality constraint when this is introduced via a Lagrange multiplier field.
It is convenient to choose a functional basis for the other constraints by defining
\begin{equation}\label{simplecons}
H_\perp[\beta] = \int\! d\sigma\,  \beta {\cal H}_\perp\, , \qquad H_\parallel[\alpha] = \int\! d\sigma\,  \alpha {\cal H}_\parallel\, , 
\end{equation}
where $\alpha$ is a 1D vector field and $\beta$ a scalar inverse-density, equivalent to a vector field in 1D. Assuming that $(\alpha,\beta)$ are smooth,  
and have compact support, one finds that 
\begin{eqnarray}\label{conPBs}
\{ H_\perp[\beta] , H_\perp[\tilde\beta] \}_{PB} &=& T^2 H_\parallel \left[ [\beta,\tilde\beta]\right] \, ,  \nonumber \\
\{ H_\parallel [\alpha] , H_\perp[\beta] \}_{PB} &=& H_\perp\left[ [\alpha,\beta]\right] \, , \nonumber \\
\{ H_\parallel [\alpha] , H_\parallel[\tilde\alpha] \}_{PB} &=& H_\parallel\left[ [\alpha,\tilde\alpha]\right] \, , 
\end{eqnarray}
where $[\cdot,\cdot]$ indicates a commutator of (1D) vector fields. As this is the same result that one finds prior to the introduction of the chirality constraint, 
we may again impose the Monge gauge conditions (\ref{Monge2}) and solve the constraints for $P_\mu$, but this now yields (for the same sign choice as before) 
the $T$-independent result
\begin{equation}\label{Pmu2}
P_0= -(\varphi')^2\, , \qquad P_1 = - (\varphi')^2 \, . 
\end{equation}
This gives $P^0=(\varphi')^2$, and the Monge gauge action becomes the FJ
chiral boson action (\ref{FJact2}).  

We have now found a covariant action for the FJ chiral boson equation but, as things stand, this has been accomplished at the cost of introducing the 
dimensionful constant $T$. Moreover, this $T$-dependence cannot be removed from the covariant action by taking $T\to\infty$ because the PB algebra of  its 
constraint functions is singular in this limit.  However, one can take the $T\to0$ limit; in this case
\begin{equation}\label{simple}
{\cal H}_\perp = \frac12 P^2\, , \qquad {\cal H}_\parallel = (X^\mu)'P_\mu + (\varphi')^2\, . 
\end{equation}
Apart from the fact that the string worldsheet fills the 2D spacetime, implying an absence of any dynamics if $\varphi'\equiv 0$ the constraints
are those of the  null, or tensionless,  string \cite{Schild:1976vq} modified by the $(\varphi')^2$ term in ${\cal H}_\parallel$. This suggests a null-string dust 
\cite{Stachel:1980zs} interpretation, and hence conformal invariance, which is readily verified: the corresponding Noether charges are
\begin{equation}\label{confN} 
Q_{(k)} = -\int\! d\sigma\  k^\mu(X)  P_\mu
\end{equation}
for any 2D conformal Killing vector field $k(X)$. 

A covariant action for the FJ chiral boson has now been found via the introduction of additional fields that can be removed by 
gauge fixing, as for the PST method \cite{Pasti:1996vs} but here the Lorentz group acts only on the additional fields. Moreover, it acts as an `internal'
symmetry that becomes a 2D spacetime Lorentz symmetry only after gauge fixing; this happens because the Monge gauge condition
is Lorentz invariant only if a  Lorentz transformation is combined with a ``compensating'' worldsheet diffeomorphism.  As the Lorentz-boost Noether charge
of (\ref{boost}) is diffeomorphism invariant, we may find the Lorentz-boost Noether charge of the Monge-gauge action by substitution using 
(\ref{Monge2}) and (\ref{Pmu2}): 
\begin{equation}
L\to L[\varphi] := - \int\! d\sigma \, \xi^+ (\varphi')^2 \, ,  \qquad (\xi^\pm =t\pm \sigma)\, . 
\end{equation}
The Lorentz-boost transformation of any function $f$ of phase-space fields is given by the formula 
\begin{equation}\label{Lboost}
\delta_w f = w\left\{f, L\right\}_{PB} \, , 
\end{equation}
where $w$ is the boost parameter.  Prior to gauge fixing, $\varphi$ is inert, but after gauge fixing we get  the  result of \cite{Floreanini:1987as}:
\begin{eqnarray}
\delta_w\varphi = - w\, \xi^+ \varphi' \,  \qquad ({\rm Monge\  gauge})\, . 
\end{eqnarray}
More generally,  the Monge-gauge expression for the Noether charge (\ref{confN}) is 
\begin{equation}
Q_{(k)} = \int\! d\sigma\,  k^+(\xi) (\varphi')^2\,  \qquad (k^\pm = k^0\pm k^1), 
\end{equation}
which generates the transformation $\delta_k\varphi=  k^+\varphi'$. Ignoring surface terms, it may be verified directly that this induces a variation of the FJ action 
that is a surface term if, and only if, $\partial_- k^+=0$, as required for a conformal transformation.  Using (\ref{gensol}) we have 
$\delta_k \Phi  = k^+ \partial_+ \Phi  \equiv {\cal L}_k \Phi$,  which is the first-order variation of a 2D scalar  field under a conformal transformation. 

Although  the simplest set of constraint functions for the covariant chiral boson action (\ref{string22}) are those of (\ref{simple}), we are free to choose those of (\ref{Hamcons}) 
and this freedom allows a string-theory application. So far we have considered a chiral reduction on $S^1$ of a 3D string. Consider now the chiral reduction on $T^{16}$ of a 26D string; the result is, 
for 10D Minkowski coordinates $\{X^M; M=0,\dots,9\}$, 
\begin{equation}
S = \int\! dt\int\! d\sigma \left\{ \dot X^M P_M +  \dot{\boldsymbol{\varphi}}\cdot \boldsymbol{\varphi}' - e{\cal H}_\perp - u{\cal H}_\parallel  \right\}\, , 
\end{equation}
where  $\boldsymbol{\varphi}$ is a Euclidean $16$-vector of FJ scalar fields, and the constraint functions are
\begin{eqnarray}\label{Hamcons2}
{\cal H}_\perp &=& \frac12 \left[\eta^{MN}P_MP_N  + 2T |\boldsymbol{\varphi}'|^2 + T^2 \eta_{MN}(X^M)'(X^N)' \right]\, ,\nonumber \\
{\cal H}_\parallel &=& (X^M)'P_N + |\boldsymbol{\varphi}'|^2 \, . 
\end{eqnarray}
In the conformal gauge,  $u=0$ and $e=1/T$,  elimination of $P_M$ yields the quadratic Lagrangian density 
\begin{equation}
{\cal L} = -2T \, \partial_-X^M\partial_+X^N \eta_{MN}  + \left(\dot{\boldsymbol{\varphi}} - \boldsymbol{\varphi}'\right) \cdot \boldsymbol{\varphi}' \, . 
\end{equation}
By adding the 10-vector of anti-chiral fermions required for  $(1,0)$-supersymmetry, and then the conformal-gauge ghosts, we arrive at an anomaly-free  conformal-gauge action for the heterotic string in which the centre of mass motion is in the 10D Minkowski spacetime because the 16 FJ chiral bosons have no zero modes.  

We turn now to the higher-dimensional chiral generalizations of the  FJ theory. We start from a $(4k+2)$-dimensional  Minkowski spacetime with Minkowski coordinates 
$(t,\boldsymbol{\sigma}=\{\sigma^i; i=1,\dots,4k+1\})$. We then introduce a $2k$-form potential $A(t)$ on the constant-time hypersurfaces, from which we 
form the gauge-invariant  antisymmetric tensor density with components
\begin{equation}
B^{i_1\dots i_{2k}}= \frac{1}{(2k)!} \varepsilon^{i_1\cdots i_{2k} j \ell_1\cdots \ell_{2k}} \partial_j A_{\ell_1\cdots \ell_{2k}}\, .
\end{equation}
Next, we propose an action of the form
\begin{equation}\label{actk}
S[A] = \int\! dt\! \int \! d\boldsymbol{\sigma} \left\{ \frac{1}{(2k)!}\dot A_{i_1 \cdots i_{2k}} B^{i_1\cdots i_{2k}} - {\cal H}\right\}, 
\end{equation} 
where ${\cal H}$ is some  gauge-invariant and rotationally-invariant Hamiltonian density; for a ``chiral'' electrodynamics we expect it to 
be a function of $B$ only. Consider the choice
\begin{equation}\label{hamdens}
{\cal H}= \sqrt{T^2 + 2T|B|^2 + |B\times B|^2}  -T\, , 
\end{equation}
where $T$ is a constant (with dimensions of  $(4k+1)$-brane tension). We use the notation
\begin{equation}
(B\times B)_i := \frac{1}{[(2k)!]^2} \varepsilon_{i j_1\cdots j_{2k} \ell_1 \cdots \ell_{2k}} B^{j_1\cdots j_{2k}} B^{\ell_1\cdots\ell_{2k}} \, ,  
\end{equation}
and $|\cdot|$ indicates the Euclidean norm: 
\begin{eqnarray}
|B|^2  &=& \frac{1}{(2k)!} \delta_{i_ij_1} \cdots \delta_{i_{2k} j_{2k}} B^{i_1\cdots i_{2k}} B^{j_1\cdots j_{2k}} \, , \nonumber \\
|B\times B| &=& \delta ^{ij}(B\times B)_i (B\times B)_j  \, . 
\end{eqnarray}
The $k=0$ case of (\ref{actk}) is the FJ action, irrespective of the value of $T$.  To see this write $A=\varphi$ for the $0$-form potential, 
so that $B=\varphi'$; then  ${\cal H}= (\varphi')^2$.  

For $k\ge1$ the Hamiltonian density (\ref{hamdens}) is $T$-dependent.  In the  $T\to\infty$ limit it is quadratic in $B$ and the action is the Henneaux-Teitelboim action for free chiral $2k$-form 
electrodynamics \cite{Henneaux:1988gg}. For finite $T$ there are interactions but these simplify in the $T\to0$ limit:
\begin{equation}\label{conflim}
{\cal H}|_{T=0} = |B\times B|\, . 
\end{equation}
This is homogeneous of degree $2$ in $B$ but not quadratic, so it defines a (conformally invariant) interacting theory. The $k=1$ case was found by Gibbons and West \cite{Gibbons:2000ck} from 
the M5-brane phase-space action of   \cite{Bergshoeff:1998vx} by a  limit analogous to that introduced by Bialynicki-Birula in the context of 
Born-Infeld electrodynamics \cite{BialynickiBirula:1984tx}.  It was recently rediscovered by the author  \cite{Townsend:2019ils}, and  interpreted as an infra-red limit of the  
chiral $2$-form electrodynamics on a static planar M5-brane in the AdS$_7\times S^4$ vacuum of M-theory; this was inspired by a similar interpretation
of Bialynicki-Birula's non-linear conformal electrodynamics as an infra-red limit of the Born-Infeld electrodynamics on a $D3$-brane 
in AdS$_5\times S^5$ \cite{Mezincescu:2019vxk}.

As we have seen here for $k=0$, and as shown for $k=1$ in \cite{Townsend:2019ils}, the action (\ref{actk}) is Lorentz invariant for the Hamiltonian density of 
(\ref{hamdens}) because it is the gauge-fixed version of a reparametrization invariant action that is manifestly Lorentz invariant.  We now investigate
whether this remains true for $k>1$. The first step is to consider an action of the form
\begin{eqnarray}\label{kcov}
S&=& \int\! dt \int\! d\boldsymbol{\sigma} \Big\{ \dot X^\mu P_\mu -\frac{1}{(2k)!}\dot A_{i_1 \cdots i_{2k}} B^{i_1\cdots i_{2k}} \nonumber \\
&& \qquad \qquad  \qquad - \, e{\cal H}_\perp - u^i {\cal H}_i\Big\}\, , 
\end{eqnarray}
for $\mu=0,1, \dots, 4k+2$, and 
\begin{eqnarray}\label{gencons}
{\cal H}_\perp &=& \frac12\left[ \mathbb{P}^2 + 2T |B^2| + T^2\det h \right] \, , \nonumber\\
{\cal H}_i &=& \partial_i X^\mu P_\mu + (B\times B)_i\, , 
\end{eqnarray}
where $h$ is the induced space metric ($h_{ij} = \partial_i X^\mu \partial_j X^\nu \eta_{\mu\nu}$ for $i,j=1, \dots, 4k+1$) also used here in $|B^2|$, and 
\begin{equation}
\mathbb{P}_\mu = P_\mu - T C_\mu\, , 
\end{equation}
where, for $d= 4k+1$, 
\begin{equation}
C_\mu= \frac{1}{d!} \varepsilon_{\mu \nu_1\dots \nu_d}\varepsilon^{i_1\dots i_d} \partial_{i_1} X^{\nu_1} \cdots\partial_{i_d} X^{\nu_d} \, . 
\end{equation}
The replacement $P_\mu\to \mathbb{P}_\mu$ serves the same purpose  as before: subtraction of $T$ from the Monge-gauge energy density, as can be verified using the identities $\partial_i X^\mu C_\mu\equiv 0$ and $C^2 \equiv - \det h$.  

The Monge gauge choice $(X^0, {\bf X}) = (t,\boldsymbol{\sigma})$ takes the covariant action of (\ref{kcov}) to the action of 
(\ref{actk}) with Hamiltonian density (\ref{hamdens}) for any $k$, but the validity of this step depends on the first-class property 
of the constraints because they will not otherwise generate the required gauge invariances. Thus, the problem of finding a
Hamiltonian density ${\cal H}$ that ensures a Lorentz invariant action,  which has been addressed by a very 
different method in \cite{Buratti:2019guq}, is here transferred to the problem of finding first-class constraint functions for the 
covariant action. It is again convenient to choose a functional basis for them by defining 
\begin{equation}
H_\perp[\beta] = \int\! d\boldsymbol{\sigma}\,  \beta {\cal H}_\perp\, , \qquad H_\parallel[\boldsymbol{\alpha}] = \int\! d\boldsymbol{\sigma}\,  \alpha^i {\cal H}_i \, , 
\end{equation}
for vector field $\boldsymbol{\alpha}$ and scalar inverse-density $\beta$ on the $(4k+1)$-space. The relevant PB relations are 
\begin{eqnarray}
\left\{X^\mu(\boldsymbol{\sigma}), P_\nu(\boldsymbol{\varsigma}) \right\}_{PB} &=& \delta^\mu_\nu \, \delta(\boldsymbol{\sigma} - \boldsymbol{\varsigma}) \, , \\
\left\{A_{i_1\dots i_n}(\boldsymbol{\sigma}), B^{j_1\dots j_n}(\boldsymbol{\varsigma})\right\}_{PB} &=&\! \frac{1}{2n!}\delta_{[i_1}^{j_1} \cdots \delta_{i_n]}^{j_n} \, 
\delta(\boldsymbol{\sigma} - \boldsymbol{\varsigma}) \nonumber 
\end{eqnarray}
for $n=2k$. 
The second line generalizes to arbitrary $k$ the relation $\{\varphi(\sigma),\varphi'(\varsigma)\}_{PB} = \frac12 \delta(\sigma-\varsigma)$, which follows directly from the 
canonical PB relations of  (\ref{PBrels}). 

Using these PB relations we may compute the PBs of the constraint functions (\ref{gencons}); for $T=0$ one finds that the only non-zero PB relations are
\begin{eqnarray}\label{conPBs2}
\{ H_\parallel [\boldsymbol{\alpha}] , H_\perp[\beta] \}_{PB} &=& H_\perp\left[{\cal L}_{\boldsymbol{\alpha}}\beta\right] \, , \nonumber \\
\{ H_\parallel [\boldsymbol{\alpha}] , H_\parallel [\boldsymbol{\tilde\alpha}]  \}_{PB} &=& H_\parallel \left[ [\boldsymbol{\alpha},\boldsymbol{\tilde\alpha}]\right] \, , 
\end{eqnarray}
where ${\cal L}_{\boldsymbol{\alpha}} \beta$ is the Lie derivative of $\beta$ with respect to the vector field $\boldsymbol{\alpha}$.
This is exactly what one finds for the null $(4k+1)$-brane,  so we may impose the Monge gauge to arrive at the action (\ref{actk}) with  Hamiltonian density (\ref{conflim}).
This is the promised class of  conformal  chiral $2k$-form electrodynamics that includes, as the $k=0$ case, the FJ chiral boson theory. 

For $k=1$, the action (\ref{actk}) is Lorentz invariant for the more general $T$-dependent Hamiltonian density of (\ref{hamdens}). In this case  
one finds that 
\begin{eqnarray}\label{perps}
\{ H_\perp[\beta] , H_\perp[\tilde\beta] \}_{PB} &=& H_\parallel [\boldsymbol{\alpha}(\beta,\tilde\beta)] \, , 
\end{eqnarray}
where 
\begin{equation}
\alpha^i(\beta,\tilde\beta) =  \left[T^2\det h\, h^{ij} + 2T[B^2]^{ij}\right] (\beta{\buildrel \leftrightarrow \over \partial}_j \tilde\beta)\, , 
\end{equation}
with $[B^2]^{ij} = B^{ik}B^{jl}h_{kl}$. The right-hand side of (\ref{perps}) apparently includes a quartic term in $B$ that cannot appear on
the left-hand side but the identity $[B^2]^{ij}(B\times B)_j \equiv 0$ ensures that it is absent \cite{Townsend:2019ils}.  This identity fails for $k>1$,  
so the Hamiltonian density of (\ref{hamdens}) does not define a Lorentz invariant theory for $k>1$ unless $T=0$.  There may exist a modified choice of ${\cal H}$ that
overcomes this restriction, but  the $k=2$ example provided by  IIB supergravity \cite{Schwarz:1983wa} suggests that  non-conformal interactions for $k>1$ must be gravitational. 
\bigskip

\noindent
{\bf Note added}: Conformal $n$-form electrodynamics in spacetime dimension $2(n+1)$ for any $n$, generalizing the $n=1$ case of Bialynicki-Birula \cite{BialynickiBirula:1984tx}, 
has been studied previously by Chru\' sci\' nski \cite{Chruscinski:2000zm}, but without the imposition of a chirality constraint for even $n$. 
\smallskip

\noindent\textbf{Acknowledgements}: I am grateful to Igor Bandos, Gary Gibbons and Dmitri Sorokin for bringing to my attention relevant earlier work,  and for other 
helpful comments on an earlier version of this paper.  This work was partially supported by STFC consolidated grant ST/P000681/1

\end{document}